# A Conceptual Exploration of Generative AI-Induced Cognitive Dissonance and its Emergence in University-Level Academic Writing

Carl Errol Seran[1,2,3], Myles Joshua Toledo Tan[1,2,4,5,6,7,8,*], Hezerul Abdul Karim[9,*], Nouar AlDahoul[10]

[1] Biology Program, College of Arts and Sciences, University of St. La Salle, Bacolod, Philippines
[2] Department of Natural Sciences, College of Arts and Sciences, University of St. La Salle, Bacolod, Philippines
[3] Faculty of Education, University of the Philippines Open University, Los Baños, Philippines
[4] Department of Electrical and Computer Engineering, Herbert Wertheim College of Engineering, University of Florida, Gainesville, FL, United States
[5] Department of Epidemiology, College of Public Health and Health Professions and College of Medicine, University of Florida, Gainesville, FL, United States
[6] Department of Chemical Engineering, College of Engineering and Technology, University of St. La Salle, Bacolod, Philippines
[7] Department of Electronics Engineering, College of Engineering and Technology, University of St. La Salle, Bacolod, Philippines
[8] Yo-Vivo Corporation, Bacolod, Philippines
[9] Faculty of Engineering, Multimedia University, Persiaran Multimedia, Cyberjaya, Malaysia
[10] Department of Computer Science, Division of Science, New York University Abu Dhabi, Abu Dhabi, United Arab Emirates

* Correspondence: Myles Joshua Toledo Tan, mylesjoshua.tan@medicine.ufl.edu; Hezerul Abdul Karim, hezerul@mmu.edu.my

## Introduction

University-level academic writing is a form of scholarly communication that demands precision, clarity and adherence to established writing conventions. It is a critical skill that is widely recognized as a key component of science education, which is best developed through practice (Moskovitz & Kellogg, 2011). The growing complexity and demands of academic writing in the university presents a multitude of challenges that stem from both educational and systemic issues. Science students often perceive writing as a daunting task due to insufficient instruction between technical and creative writing during their education (Strive, 2024). This aversion is compounded with the challenge of writing articles that requires clear, logical narratives, especially for non-native English speakers (Nazaroff, 2011). Furthermore, the perceived marginal role of writing in research and teaching adds to this challenge. Students and even educators struggle with mastering the disciplinary discourses that are essential for knowledge construction and career advancement (Hyland, 2013). Moreover, lesser attention to grammar instruction, and focus on the writing process often lead to poorly refined scientific documents. Many instructors believe that this is the sole responsibility of the English department (Alley, 2024). The emergence of Generative Artificial Intelligence (GenAI) has also led some students to undervalue the importance of developing writing skills, assuming that technology could compensate for their deficiency.

GenAI tools, such as large language models, e.g. ChatGPT[TM] by OpenAI, Gemini[TM] by Google, and Claude[TM] by Anthropic, provide a practical solution to organize and simplify writing-intensive tasks, such as article reviews, lab reports, and experimental research. These tools reduce cognitive load by allowing students to focus on critical analysis and interpretation, instead of becoming overly focused on the technical aspects of writing (Essel et al., 2023). It also guides university students to adhere to language precision and logical structure of scientific communication (Olatunbosun & Nwankwo, 2024; Preiksaitis & Rose, 2023). However, while it is highly beneficial for reducing cognitive workload and productivity, its integration often creates a blurred boundary between perceptions of ownership and intellectual effort (Amoozadeh et al., 2023). This ambiguity gives rise to *cognitive dissonance*, a term deeply rooted in classic social psychology introduced by Festinger (1957). Cognitive dissonance (CD) is the inconsistency in thoughts, actions or behavior. When we hold two opposing beliefs or ideas, it creates tension or discomfort (Hilberg, 2017). Through this experience, we resolve the tension by either changing our thoughts or behaviors, adding a new thought, or rationalizing the inconsistencies (Oxoby & Smith, 2014). CD becomes increasingly pronounced when students struggle with reconciling their reliance on GenAI tools with academic values of originality and integrity.

To date, no consensus has been made regarding the manifestation of cognitive dissonance in the integration of GenAI in university-level academic writing. Here, we provide a novel perspective on how cognitive dissonance emerges with the integration of GenAI in university-level academic writing. We posit that the integration of GenAI in university-level academic writing could potentially initiate or amplify existing triggers of CD.

**Understanding Cognitive Dissonance in Traditional Academic Writing**

CD is a long-standing issue in academic writing that affects both educational processes and outcomes (McGrath, 2017). This concept is operationally defined with shared shortcomings, across different studies and contexts. For the purpose of clarity, CD is operationalized as a state of psychological discomfort or tension, and a trigger of a cognitive dissonance state (CDS) (Vaidis & Bran, 2019). This tension often manifests when students face conflicting demands between their academic ideals and practical constraints. In scientific writing, it may manifest as ethical dilemmas, such as pressures to publish while adhering to strict standards of originality and rigor (Schrems & Upham, 2020). It can also arise from the discrepancies between the ideals of producing quality work and the practical challenges of managing limited time and resources. Furthermore, students may experience tension when balancing the need for clarity and simplicity, and the complexity of the subject matter. When not handled, this often leads to dishonesty (Stephens, 2017). This dissonance may result in coping strategies like selectively focusing on information that aligns with their existing beliefs (Metzger et al., 2015). This state of dissonance continually pushes students to recalibrate their beliefs and values to produce quality work while upholding academic integrity. CD in scientific writing may be viewed as both a psychological stress and a catalyst for positive change. While it is a recognized challenge in traditional scientific writing, the emergence of GenAI tools has introduced new dimensions to this phenomenon.

**The Dual Impact of GenAI in University-Level Academic Writing**

University students may regard the duality of GenAI in academic writing as both beneficial and challenging. Many students appreciate the efficiency and support that GenAI tools, like ChatGPT™, provide in generating outlines, summarizing research, and refining language used (Dergaa et al., 2023). Additionally, GenAI supports students in managing dense and complex experimental data, making it easier to present results clearly and accurately (Gao et al., 2023). However, the psychological discomfort associated with using GenAI in academic writing blurs the boundary between human effort and AI assistance. There is a growing concern regarding the potential for over-reliance on AI, which may lead to questioning one's own writing abilities and originality. The systematic review of 14 articles by Zhai et al. (2024) suggests that ethical concerns related to AI lead to over-reliance, which affects students' cognitive abilities as they prioritize faster, optimal solutions over slower, more practical approaches. The ease of generating text with AI leads to a superficial understanding of the writing process. This preference emphasizes why users often rely on efficient cognitive shortcuts, or heuristics, despite the ethical challenges posed by AI technologies. Students initially recognize the benefits of AI in enhancing productivity and providing support. It also crosses the line where they feel guilty or anxious about relying on technology to produce work that is expected to reflect their own knowledge and creativity (Wang, 2024). This dependency creates a cycle where students increasingly rely on AI-generated outputs. If not addressed, this could eventually erode their confidence in producing original work.

While CD is well-studied in traditional contexts such as the occurrence of academic dishonesty, and ethical dilemmas, the concept of GenAI-induced CD in academic writing is still underexplored. The integration of GenAI tools in university-level academic writing challenges traditional notions of originality, effort, and academic integrity, leading to the manifestation of CD.

**GenAI-induced Cognitive Dissonance: A Hypothetical Construct**

*GenAI as a Potentially Trigger for Cognitive Dissonance*

A CD trigger is any external factor that disrupts an individual's beliefs, attitudes, or values, creating psychological tension. The efficiency and convenience of GenAI often conflict with academic principles like originality, effort, and integrity, leaving students to confront ethical dilemmas in its use. This tension creates a fertile ground for CD, as students grapple with the ethical implications of relying on AI for academic tasks. Empirical studies by Chan (2024), Ju (2023), and Playfoot and Quigley (2024) provide robust evidence of how GenAI serves as a significant trigger for cognitive dissonance, highlighting the psychological and ethical dilemmas it introduces.

Chan's (2024) study, which surveyed 393 university students, provides critical insights into the phenomenon of "AIgiarism," a term describing plagiarism facilitated by AI tools. The findings reveal that while students strongly disapproved of directly submitting AI-generated content, they expressed significant ambivalence toward more nuanced uses of GenAI, such as paraphrasing and idea generation. This ambivalence reflects a broader tension between the practical benefits of AI and the ethical demands of academic integrity. Notably, the study found that CD induced by GenAI was prevalent across diverse academic disciplines, including engineering and the arts, suggesting that the psychological impact of AI is not confined to specific fields but is instead a widespread phenomenon. Building on these insights, Ju's (2023) experimental study involving 32 participants revealed that excessive reliance on GenAI tools correlates with a measurable decline in academic performance. The study found that participants who heavily depended on AI for writing tasks experienced a 25% reduction in writing accuracy and comprehension compared to those who used AI sparingly. This decline was not merely academic; it also heightened CD among students, who reported a noticeable drop in confidence and expressed doubts about the authenticity of their work. Ju's research highlights how GenAI undermines the principle of intellectual ownership, a foundational pillar of academic values, thereby intensifying psychological tension. These findings suggest that the dissonance induced by GenAI transcends ethical dilemmas, significantly affecting students' self-efficacy and their sense of academic identity. Further supporting this perspective, Playfoot and Quigley's (2024) cross-disciplinary survey of 467 students explored the adoption of AI tools, e.g., ChatGPT$^{TM}$, for academic tasks, where participants expressed ambivalence toward ethical boundaries in AI-assisted tasks like idea-generation or paraphrasing. In this study, 68% recognized that GenAI use conflicted with academic integrity, while 52% prioritized convenience and perceived low detection risks. This contradiction between practicality and integrity exemplifies CD, as students navigate competing priorities: the efficiency of GenAI versus values like originality and effort.

*GenAI as a Potential Amplifier of Pre-existing Tensions*

An *amplifier* refers to an external factor that heightens existing psychological conflicts. In academic writing, GenAI intensifies ongoing struggles, such as balancing precision with creativity, maintaining originality, and adhering to ethical standards. While improving efficiency, it can deepen dependencies and blur the boundaries of authorship, further widening the gap between academic ideals and practical outcomes. Empirical evidence demonstrates how these dependencies amplify CD, as users wrestle with conflicting priorities between AI-driven convenience and academic integrity.

The role of GenAI as an amplifier of CD in academic writing is well-documented. Zhai et al. (2024), in their systematic review of 14 studies involving 1,200 university students, found that GenAI magnifies pre-existing tensions related to academic skill development and engagement. A majority of students reported prioritizing efficiency over deep cognitive engagement, leading to greater uncertainty about their learning progress. Similarly, Ironsi and Ironsi's (2024) study of 150 university students revealed that while 70% believed GenAI improved the clarity of their writing, it did not significantly enhance their overall writing skills. This gap between perceived benefits and actual skills development created psychological tension, with 61% expressing frustration about their inability to improve independently while relying heavily on GenAI. Further supporting these findings, Hutson's (2024) study of 200 STEM students highlighted how the convenience of GenAI tools exacerbated reliance on cognitive shortcuts, with 45% of participants questioning their ability to independently complete academic tasks after repeatedly using GenAI for initial drafts. Collectively, these studies illustrate that GenAI does not merely introduce new conflicts but amplifies pre-existing struggles, intensifying students' internal tensions over academic integrity, skills mastery, and self-efficacy.

GenAI tools such as ChatGPT have emerged as a potential catalyst of CD in university-level academic writing. The empirical evidence provided, spanning surveys, experimental studies, and systematic reviews, robustly support the argument that GenAI both acts as a potential trigger and amplifier of CD. This construct aligns with Festinger's framework. It creates contradictions between the students' academic values and their reliance on AI-driven efficiency. Festinger's theory posits that dissonance arises when an individual's actions conflict with their self-concept or deeply held beliefs. In academic settings, students are taught and encouraged to value originality, critical thinking, and intellectual ownership. However, GenAI tools disrupt these values by offering shortcuts that prioritize efficiency over effort. For example, heavy reliance on GenAI undermines the academic principle of purposeful learning. It creates a tension between the desire for efficiency and the guilt of bypassing skills development. These contradictions generate psychological discomfort, compelling students to either justify their use of GenAI ("Everyone does it.") or reassess their academic values — both hallmark responses to CD. The role of GenAI as a potential trigger and amplifier share the same construct of tension. It illustrates how GenAI introduces ambiguity between what is ethical and what is convenient. This ambivalence leads to a dissonant state of struggling to reconcile convenience with integrity. Festinger's theory predicts such ambivalence as a symptom of unresolved dissonance. These predictions are in the form of psychological discomfort, behavioral justifications, value reassessment, such are evident across studies.

The relationship between GenAI, cognitive dissonance, and behavioral outcomes in academic settings is visually summarized in Figure 1, which illustrates how GenAI acts as both a trigger and amplifier of CD, ultimately shaping students' academic behaviors and perceptions.

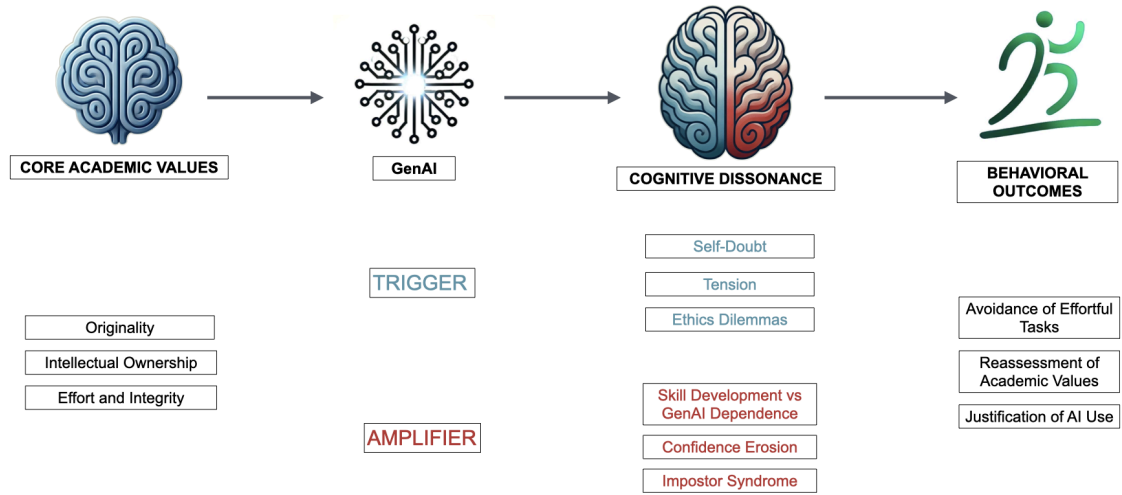

**Figure 1.** Hypothetical construct of GenAI-induced cognitive dissonance (CD). GenAI is both a trigger and an amplifier of CD in academic writing. The figure illustrates how the integration of GenAI disrupts core academic values, leading to CD, which manifests through self-doubt, ethical dilemmas, and confidence erosion, among others. These psychological tensions influence behavioral outcomes, including avoidance of effortful tasks, reassessment of academic values, and justification of AI use. The process is further amplified by increasing dependence on GenAI, which exacerbates existing struggles related to skills development and academic integrity.

**Addressing CD in University-Level Academic Writing**

The integration of GenAI into university-level academic writing has introduced new ethical and psychological challenges, particularly CD. This psychological conflict is not necessarily new with the adaptation of new technologies. However, the emergence of conflict in the case of GenAI stems from the challenge of addressing authentic human creativity, which has not been seen with past technologies. This requires proactive strategies that empower students not just to think critically but also to build their confidence in presenting their work as their own. To build students' confidence in ownership, a multifaceted approach that emphasizes transparency, education, and adherence to ethical guidelines should be implemented.

*AI Literacy Programs*

Engaging with GenAI technologies in education is still relatively a new experience for universities (Alasadi & Baiz, 2023), particularly in developing countries like the Philippines. Ethical conflicts, diminished confidence, and superficial engagement are more likely to be experienced in the absence of evaluation, validation, and responsible integration of AI-generated content. AI literacy programs can provide the critical knowledge needed to define the boundaries of human output and AI-generated work. To reduce the dissonance experienced by students, it is important to clearly define the role of GenAI, including students' engagement with it and the transparency of its utilization. Practices such as cross-checking AI-generated summaries against original research data can help reconcile the tension between convenience and academic integrity. When students recognize the limitations of GenAI in the writing process, they are likely to feel more confident in their cognitive efforts.

### *Reflective Pedagogy*

Reflective practices can help students reconcile their reliance on GenAI with their academic values and personal learning goals. These practices encourage them to examine their beliefs about academic integrity and their personal learning approaches. Practical suggestions include reflective thinking exercises on their views of academic integrity and self-assessment of their confidence in submitting assignments, such as lab reports. Through this process, students can develop a deeper understanding of their role in the writing process and address feelings of diminished ownership. This can enhance their motivation to improve cognitive and metacognitive strategies, promoting self-regulated learning (Zhai et al., 2023).

Educators should encourage prompt peer or instructor reviews of GenAI-integrated work, where feedback clearly distinguishes the role of the GenAI tool from the student's critical contributions. This approach helps students perceive GenAI as a collaborative tool that complements human effort, fostering a balanced and thoughtful perspective on its use.

### *Establishment of Transparency Standards on Ethical GenAI Integration in Critical Thinking and Creative Works*

Universities must establish clear guidelines for the transparent use of GenAI in academic writing. Setting ethical boundaries for its integration helps eliminate ambiguity and ensures that students can use AI appropriately while upholding academic values, such as originality and intellectual engagement. As suggested by Tan and Maravilla (2024), ethical guidelines and policies should be grounded in constructivist learning frameworks to ensure that GenAI use aligns with educational values and promotes authentic learning.

One recommended practice is requiring students to declare their use of GenAI in their academic works. This could reassure students that their use of GenAI aligns with institutional values, alleviating guilt or anxiety about potential misuse. Additionally, institutions should clearly define acceptable and unacceptable uses of GenAI in academic writing. For instance, they may specify that GenAI can be used for language refinement but not for generating experimental interpretations or original arguments. Requiring students to disclose AI-generated content, such as identifying AI-assisted sections in appendices or footnotes, can further promote transparency. Establishing clear academic policies will reduce confusion and empower students to navigate the ethical boundaries of GenAI use with confidence in their academic writing.

### *Redesigning Writing Tasks for Discipline-Specific Challenges*

Academic writing demands precision and critical thinking, making it essential to redesign discipline-specific tasks to emphasize key skills. For instance, students can generate an initial draft using GenAI but should be required to manually refine and expand specific sections of their work, such as the methodology, discussion, or conclusion. This approach encourages students to engage critically with their content while addressing concerns about over-reliance on GenAI. Additionally, students can compare AI-generated analyses of experimental data with their own interpretations. This practice could strengthen their confidence in their analytical skills and reinforce the idea that AI is a tool to support, rather than replace, their expertise.

**Conclusion**

The integration of GenAI into academic writing is not merely a technological shift but also a psychological challenge. Students, already burdened by the demands of academic rigor, now face the additional pressure of navigating the ethical and practical implications of GenAI. As reliance on these tools grows, students encounter a paradox: the very technology designed to assist them may undermine their confidence and sense of authorship. Empirical evidence suggests that GenAI can act as both a trigger and an amplifier of CD in academic writing. A hypothetical construct of GenAI-induced CD in academic writing is proposed to highlight the need for a balanced approach — one that integrates technological advancements while safeguarding the human elements of learning and creativity.

Technological progress is inevitable. To address the tensions it creates in education, educators and institutions must reinforce critical engagement with GenAI. This involves not only teaching students how to use the technology effectively but also encouraging reflection on its ethical and intellectual implications. Promoting a culture of academic integrity and self-awareness is essential. By acknowledging students' struggles with the complexities of academic writing and the ethical integration of GenAI, we can support them in navigating these challenges. Ultimately, the goal is to ensure that GenAI serves as a complement to human effort rather than a replacement for creativity and intellectual growth.